\documentstyle[12pt]{article}
\title{Newtonian Evolution of the Weyl Tensor}
\author{G F R Ellis and P K S Dunsby\\
Department of Applied Mathematics, University of Cape Town\\
Rondebosch 7700, Cape Town, South Africa}
\begin{document}
\maketitle
\begin{abstract}
In an interesting recent paper on the growth of inhomogeneity through
the effect of gravity [1], Bertschinger and Hamilton
derive equations for the electric and magnetic parts of the Weyl
tensor for cold dust for both General Relativity and Newtonian
theory. Their conclusion is that both in General Relativity and
in Newtonian theory, in general the magnetic part of the Weyl
tensor does not vanish, implying that the Lagrangian evolution of
the fluid is not local. We show here that the `Newtonian' theory
discussed by them is in fact not Newtonian theory {\it per se},
but rather a plausible relativistic generalisation of Newtonian
theory. Newtonian cosmology itself is highly non-local
irrespective of the behaviour of the magnetic part of the Weyl
tensor; in this respect the Bertschinger-Hamilton generalisation
is a better theory.
\end{abstract}
\newpage
%%%%%%%%%%%%%%%%%%%%%%%%%%%%%%%%%%%%%%%%%%%%%%%%%%%%%%%%%%%%%%%%%%%%%%%%
\section{The context}
The question at issue [1] is the relation between the fully covariant
analysis of fluid flows in general relativity [2-5] as compared with that
of Newtonian gravitational theory, and their
implications for the study of inhomogeneities in cosmology ([6-9]
and references therein). The particular point that has generated
interest has been the recent discovery of a family of General
Relativity exact solutions -- those with vanishing magnetic part
of the Weyl tensor -- where neighbouring flow lines evolve
independently of each other, in the sense that this evolution
depends only on the local fluid variables [10-14]. Actually it
has been known for a long time that this is true for spherically
symmetric solutions, but the solutions now under consideration
include the relativistic version of the Zeldovich theory
of gravitational collapse [15], however predicting filamentary
gravitational collapse rather than sheetlike gravitational
collapse as in the Zeldovich analysis. It has been suggested that
this difference arises because Newtonian theory is essentially
non-local, and that the relativistic theory should have a non-zero
magnetic part of the Weyl tensor to give corresponding non-local
physics. The issue then arises, does the magnetic part of the
Weyl tensor necessarily vanish in the Newtonian limit (as claimed
in [4]), and are the Lagrangian evolution equations in that
limit local in general? According to [1], the answers to both
questions are no.\\

However this discussion has not referred back to the analyses of
Newtonian cosmology by Heckmann and Schucking [16-18], where it
is shown firstly that {\it no}\, Newtonian cosmology -- based on a
potential and Poisson equation -- is possible without some
extension of Newtonian theory as normally understood; and that
the obvious extension of Newtonian theory to the cosmological
situation is essentially non-local -- so that local physics
cannot be decoupled from instantaneous boundary conditions at
infinity. This is because Newtonian theory is a singular limit of
general relativity theory, precluding the possibility of
gravitational waves proceeding at a finite speed. It does not in an
obvious way involve an analogue of the magnetic part of the
Weyl tensor [4].\\

By contrast, the Bertschinger-Hamilton theory [1] leads to influences
propagating at a finite speed, and so their effect do not arrive
at any event instantaneously from infinity. In this sense their
theory - which allows gravitational waves -- is a better
approximation to the correct classical theory of gravity (namely,
general relativity) than is Newtonian theory proper (or more
strictly, the Heckmann-Schucking version of Newtonian theory [16-18],
with boundary conditions allowing spatially
homogeneous cosmologies). In
what follows, to save repetition of these names we shall refer to
the different theories as BH (Bertschinger-Hamilton [1]), HS
(Heckmann-Schucking [16-18]), NG (standard Newtonian
gravitational theory expressed in a potential formalism), and GR
(general relativity theory).

%%%%%%%%%%%%%%%%%%%%%%%%%%%%%%%%%%%%%%%%%%%%%%%%%%%%%%%%%%%%%%%%%%%%%%%%
\section{General Relativity Theory}
In GR applied to cosmology in a covariant manner [2-5], one decomposes
geometrical and physical quantities relative to a preferred 4-velocity
vector $u^a$ - the fundamental velocity that underlies any coherent
cosmological model [2-5]. This defines the fluid expansion
$\Theta$, shear $\sigma_{ab}$, vorticity $\omega_{ab}$, and
acceleration $a_a \equiv \dot{u}_a$. Their time derivatives
(denoted by a dot, e.g. $\dot{\Theta})$ are determined by the
matter density $\mu$ and pressure $p$ (here we restrict our
consideration to perfect fluids) together with the electric part
$E_{ab}$ of the Weyl tensor, whose derivative in turn is
determined by the magnetic part $H_{ab}$ of the Weyl tensor,
where these quantities are defined respectively from the Weyl
tensor $C_{abcd}$ by
\begin{equation}
E_{ac} = C_{abcd} u^b u^d, ~~~ H_{ac} = C_{abef} u^b
\eta^{ef}{}_{ch} u^h
\end{equation} %1
(each being a trace-free symmetric tensor that is orthogonal to
$u^a$). The quantities $E_{ab}$ and $H_{ab}$ obey
equations very similar in structure to Maxwell's equations written
relative to a general family of observers [3-5]. A particular
consequence is that $E_{ab}$ and $H_{ab}$ each obey a wave
equation, with somewhat complicated source terms (see [3] for the
case of an almost-Robertson-Walker space time).\\

In the case of interest, we specialize to `dust', that is the
pressure $p$ vanishes and consequently the fundamental flow lines
are geodesic ($a_a = \dot{u}_a = 0$). The resulting equations are
given below.

%%%%%%%%%%%%%%%%%%%%%%%%%%%%%%%%%%%%%%%%%%%%%%%%%%%%%%%%%%%%%%%%%%%%%%%%
\section{Newtonian Theory and Cosmology}
Attempts to use Newton's force law in the case of an infinitely
spread out medium, such as in a spatially homogeneous Newtonian
cosmology, are plagued by infinities (because the matter source
of the gravitational force is unbounded), so in setting up the equations of
Newtonian cosmology, an approach based on a potential $\Phi$
is preferable. \\

The potential must satisfy the Poisson equation:
\begin{equation}
\nabla^2 \Phi = - 4 \pi G \rho
\end{equation} %2
where $\rho$ is the matter density. This has to be supplemented
by boundary conditions at infinity in order to get a unique
solution; the usual ones are
\begin{equation}
\lim_{r \rightarrow \infty} \Phi = 0\,.
\end{equation} %3
We take equns. (2,3) as defining standard Newtonian gravity. Now
the problem is that these cannot give us a sensible cosmological
model, for (3) is incompatible with a homogeneous distribution of
matter [16]; in order to deal with an infinite matter distribution we
have to drop (3) and replace it by some other boundary condition
which allows the potential to diverge at infinity, as for example
in the Newtonian version of the Robertson-Walker universe
models.\\

What are suitable boundary conditions? Following HS, we define
the Newtonian analogue of $E_{ab}$ by the equation
\begin{equation}
E_{\alpha \beta} = \Phi_{,\alpha\beta} - {1 \over 3}
h_{\alpha\beta} \nabla^2\Phi ~~~\Rightarrow ~~E_{\alpha\beta} =
E_{\beta\alpha}, ~~E^\alpha{}_\alpha = 0\, .
\end{equation}%4
Then the Newtonian version of Robertson-Walker models are compatible with
the condition
\begin{equation}
\lim_{r \rightarrow \infty} E_{\alpha \beta} = 0\,,
\end{equation} %5
which is thus a weakening of condition (3) that allows one to
handle spatially homogeneous cosmological models, which
necessarily have uniform density\footnote{One also has to
generalise the idea of inertial motion to free-fall motion, in
analogy with general relativity, in order to define spatially
homogeneous Newtonian models such as the Newtonian version of the
FRW models; that issue is not of concern here.}. However this
conditions excludes the Newtonian analogs of the Kantowski-Sachs
and Bianchi spatially homogeneous but anisotropic universe models - which
are amongst the simplest generalisations of the FRW universes. Thus HS
proposed instead the condition
\begin{equation}
\lim_{r \rightarrow \infty} E_{\alpha \beta} = E_{\alpha
\beta}(t)|_\infty
\end{equation} %6
where $E_{\alpha \beta}(t)|_\infty$ are arbitrary functions of time
\footnote{in general they should depend on polar angles $\theta$
and $\phi$ also; however in the context of spatially homogeneous analogues
of the Bianchi models, they can depend only on $t$.}.
Thus the limit of the components $E_{\alpha \beta}$ at spatial
infinity may be arbitrarily prescribed as functions of time; they
then propagate in to any point with infinite speed, according to
the equation
\begin{equation}
E^\alpha{}_{\beta,\alpha} = {8\pi\over3}G  \rho_{,\alpha}
\end{equation} %7
which follows from the Poisson equation (2) and definition (4).\\

There has been some debate over these boundary conditions,
Narlikar [19] suggesting that (5) are in fact better conditions to use
for Newtonian cosmology than (6) - but thereby excluding from his
consideration that family of NT models that correspond to the
Bianchi GR solutions allowed by (6) ([17,20]). Both are of course
generalisations of the `true' Newtonian conditions (3).
Incidentally, one can ask here why one does not specify either
$\lim_{r\rightarrow\infty} \Phi = \Phi(t)|_\infty$ or $\lim_{r
\rightarrow \infty} \nabla^2 \Phi$; the reason is that the first
diverges, while the second is determined by the limiting mass
density at infinity (through the Poisson equation (2)).\\

The point that is fundamental to us here is that NT itself gives
{\it no} equation for $\Phi\,\dot{}$, and consequently can give
{\it no} equation for $E\,\dot{}_{\alpha\beta}$ (with
$E_{\alpha\beta}$ defined by (4)). This is true whether one
writes NT in Lagrangian or Eulerian coordinates. Any
specification one may obtain for these time derivatives thus {\it
either} results from positing a theory that is not NT itself, but
rather some generalisation of NT; {\it or} from positing some set
of boundary conditions, such as (5) or (6). The
latter option does not directly give any local {\it equation} for
$E\,\dot{}_{\alpha\beta}$, which is then determined non-locally by
the chosen condition at infinity plus the local matter
conservation equations and the divergence equation (7) for
$E_{\alpha\beta}$. NT itself (which obeys (3)) gives no clue as
to which of these generalised boundary conditions ((6) or its
specialisation (5)) we should choose. \\

We support the HS view that (6) rather than (5) is the better option.
In both cases, information immediately propagates in from infinity to
determine the local physical response; the fact that the information
imparted in (5) is the `null' case - there is no change in
$E_{\alpha\beta}$ at infinity -  does not change the fact we are
determining what happens locally by a choice of conditions at infinity;
and according to (6), these are arbitrarily specifiable.\\

One might ask what are the suitable conditions to use at infinity
for a Newtonian version of a perturbed FRW model, that has
statistically spatially homogeneous inhomogeneities superimposed
on a FRW background (which is necessarily spatially homogeneous
and therefore stretches to infinity). It would seem we are in
trouble here, for {\it none} of the conditions above seem
adequate to this case - because then there will be {\it no}
regular limit for $E_{\alpha\beta}$ as we go to spatial infinity
(because of the statistical fluctuations in the matter, this
quantity too may be expected to fluctuate statistically). We can
get a good description {\it either} by imposing periodic boundary
conditions, as is done in most numerical simulations - thus
avoiding the problem in the way first suggested by Einstein
through his proposal of spatially compact universe models (and
corresponding to the possibility of finding Newtonian versions of
the `small universe' idea [21]); {\it or} by insisting that the
perturbations die away outside some bounded domain, thus allowing
(5) for example as a boundary condition, at the expense of
denying the assumption of spatial homogeneity of conditions in
the universe - which is one of the central tenets of current
cosmological dogma. \\

Any use of NT to discuss the growth of perturbations in a cosmological
context should make it quite clear which of these options is adopted -
and why. It may perhaps be claimed that {\it physically} these both give
reasonable results - but that has to be shown. In our view the best
alternative would be a third option that has not so far been systematically
developed: namely, to adopt a Newtonian version of the `Finite Infinity'
proposal for the GR case [22], that is, to isolate the considered local
system by a sphere that is far enough away to be regarded as infinity
for all practical purposes but, because it is at a finite distance, can be
investigated easily and used as a surface where boundary conditions can
be imposed (and the residual influence of the outer regions on the
effectively `isolated' interior can thus be determined).

%%%%%%%%%%%%%%%%%%%%%%%%%%%%%%%%%%%%%%%%%%%%%%%%%%%%%%%%%%%%%%%%%%%%%%%%
\section{The Bertschinger-Hamilton Theory}
The BH theory is developed as follows [1]:
the continuity and Poisson equations are written in `comoving' coordinates,
with a background FRW expansion factored out; thus these equations are
written for the fractional density perturbation $\delta$ and an associated
part $\phi$ of the full potential $\Phi$. This hides a tricky gauge
problem, because `comoving' here actually means relative to a fictitious
background FRW space-time (the matter moves relative to the chosen frame);
one might suggest it would in this context be preferable to use the
Newtonian version [23] of the GR gauge-invariant and covariant theory [6-9].
In the BH approach, the Newtonian gravity vector
$\vec{g}$ is defined from the perturbative potential $\phi$ by
$g = - \vec{\nabla}\phi$, and then in turn defines the tensor $E_{ij}$
by a perturbative version of (4), written in comoving coordinates. \\

BH then develop a series of equations that are very similar to those
obtained from the linearised GR equations. Substituting $\vec{g}$ into the
continuity equation gives
\begin{equation}
\vec{\nabla}.\left({\partial a \vec{g} \over \partial \tau}\right) =
4 \pi G a^3 \vec{\nabla}.\vec{f}, ~~~\vec{f} \equiv \rho \vec{v} =
\vec{f}_{||} + \vec{f}_\perp
\end{equation} %8
where the mass current has been decomposed in the comoving frame into
longitudinal and transverse parts obeying
\begin{equation}
\vec{\nabla} \times \vec{f}_{||} = 0, ~~~ \vec{\nabla}.\vec{f}_\perp = 0
\end{equation} %9
Then the transverse mass current is replaced by a transverse vector
field $\vec{H}$ obeying
\begin{equation}
\vec{\nabla}\times \vec{H} = - 16 \pi G a^2 \vec{f}_\perp,
{}~~~\vec{\nabla}.\vec{H} = 0
\end{equation} %10
 From this in turn they define a traceless tensor
\begin{equation}
{\cal H}_{ij} \equiv - {1 \over 2} \nabla_j H_i + 2 v_k \epsilon^{kl}{}_i
\nabla_j g_l  = H_{ij} + \epsilon _{ijk} A^k
\end{equation} %11
where $H_{ij}$ is the symmetric part of ${\cal H}_{ij}$ and $A^i$ the dual
of the anti-symmetric part. Now BH perform a series of manipulations, the key
ones of which are integrating (8) and substituting to get
\begin{equation}
{d\vec{g} \over d\tau} + {\dot{a}\over a}\vec{g} = 4 \pi G a^2
\bar{\rho} \vec{v} + \vec{A}
\end{equation} %12
and then taking the spatial gradient of this equation and substituting to
get
\begin{equation}
{d E_{ij} \over d\tau} + {\dot{a} \over a} E_{ij} - \nabla_k
\epsilon^{kl}{}_{(i} H_{j)l} + \Theta E_{ij} - h_{ij} \sigma^{kl} E_{kl} -
3 \sigma^k{}_{(i} E_{j)k} - \omega^k{}_{(i} E_{j)k}
= - 4 \pi G a^2 \rho \sigma_{ij}
\end{equation} %13
At this point there is something of a non-sequitur where a GR equation for
$E_{ij}$ is introduced; taking the curl of this equation and making some
substitutions BH obtain the time-derivative equation for $H_{ij}$:
\begin{equation}
{d H_{ij} \over d\tau} + {\dot{a} \over a} H_{ij} + \nabla_k
\epsilon^{kl}{}_{(i} E_{j)l} + \Theta H_{ij} - h_{ij} \sigma^{kl} H_{kl} -
3 \sigma^k{}_{(i} H_{j)k} - \omega^k{}_{(i} H_{j)k} = 0.
\end{equation} %14
Together with the divergence equations for $E_{ij}$ and $H_{ij}$, these give
the set of Maxwell-like equations for $E_{ij}$ and $H_{ij}$ that occur
in the linearised version of GR.\\

At this point it is quite clear that we have a theory that is something other
than NG, because (1) NG cannot determine the time evolution of $E_{ab}$ by
an equation like (13), as discussed in the previous section,
and (2) if we substitute (14) into the time derivative of (13) we
get a wave equation for $E_{ab}$, showing the possibility of gravitational
waves (with speed unity in the chosen units) in the BH theory (as in the
linearised GR theory, see [3]). However it is also clear on comparison with
the GR equations (see [1]) that the BH theory is a good generalisation of
Newtonian theory, in that it gives a good approximation to the GR equations.\\

How has it come about that the BH theory has produced a time evolution
equation for $E_{ij}$, and hence (with the time evolution equation for
$H_{ij}$) a wave equation for $E_{ij}$? The first point is
that, in going from (8) to (12), the solutions to the equations are
very under-determined, for example, there is considerable arbitrariness
in $\vec{g}$. Again in the series of manipulations that leads from
$\vec{f}$ to $\vec{H}$ to $H_{ij}$ and $A^k$, there is considerable
arbitrariness in these quantities if one allows for the general possible
solutions. In fact if one traces what happens in these equations, actually
(from the Newtonian viewpoint), {\it the tensor $H_{ij}$ is arbitrarily
specifiable}. This is because $\vec{f}$ determines only the $curl$ of
$\vec{H}$ (by (10)); on integrating to determine $\vec{H}$, the trace-free
symmetric part of $\vec{H}$ is (at least locally) arbitrarily specifiable;
but this is just $H_{ij}$. More precisely, the integrability equations
for $H_i$ to exist, given $H_{ij}$, are just the `divergence' equations
for $H_{ij}$ ((43) in BH); so $H_{ij}$ can be chosen as an arbitrary
solution of these spatial equations, allowing arbitrary time evolution.
Furthermore because of the arbitrary functions that occur in integrating
(8), there is also considerable arbitrariness in $A^i$.\\

How then do we get an equation ((13) here, (49) in BH) that gives the
time derivative of $E_{ij}$ in terms of $H_{ij}$? Our claim is that
in fact in this equation, {\it both $dE_{ij}/dt$  and $H_{ij}$ are
arbitrarily specifiable in NT}. From this viewpoint, $E_{ij}$ is arbitrarily
specifiable as a function of time, and (13) then tells us what $H_{ij}$ has
to be, given the definition (11) (or one can run this the other way: choose
$H_{ij}$ as you like, and (13) then determines $dE_{ij}/d\tau$). Hence this
equation does not actually determine the time evolution of $E_{ij}$ - for
there is no Newtonian equation for the time evolution of $H_{ij}$ (in a
truly NT approach, we might {\it define} $H_{ij}$ by (13), and accept
whatever identities follow from this definition; none then determine the
time evolution of the system [17]).\\

What then of the BH equations that determine the time derivative of $H_{ij}$
(and hence lead to the wave equation already commented on)? Well, in order
to derive them BH introduce their equation (52), which follows from GR rather
than NT. Thus it is here that they close the equations in causal terms -
by importing relations that do not in fact follow from NT. The time
derivative equation (14) for $H_{ij}$ (their equation (55)) then follows, and
hence the existence of gravitational waves (and the implied speed
of travel of those waves).\\

In summary, the $E_{ij}$ evolution equations arise from use of definitions
that bring the NT theory into the form of the GR equations, but do
not in fact determine the time evolution uniquely, as long
as we remain within the confines of NG (rather these equations define the
variable $H_{ij}$ which can absorb any chosen time evolution).
This does not prove wrong the set of equations given, which in fact allow
rather arbitrary solutions; however the appearance of uniqueness
is misleading, resulting from non-uniqueness of
solutions to the equations presented. One can obtain uniqueness by use of
special rather than general solutions to the equations (just as (5)
is a special case of (6)), but this is an arbitrary restriction on the
allowed solutions of the theory. The $H_{ij}$ evolution equation does not
result directly from NT at all, but rather is imported from GR. \\

Thus the BH theory
becomes determinate locally (and in good accord with linearised GR) only
by introducing extra equations which do not in fact follow from NG. The
result is a theory that is {\it a better approximation to GR than NT is},
\, in particular because it allows gravitational waves. This theory is more
local than Newtonian theory proper, as it has Cauchy development properties
like GR. It also has the advantage of having a quantity $H_{ij}$ that
corresponds in the appropriate way (in terms of its role in the equations)
to the magnetic part of the Weyl tensor in GR (NG theory proper has no such
tensor [4]).

%%%%%%%%%%%%%%%%%%%%%%%%%%%%%%%%%%%%%%%%%%%%%%%%%%%%%%%%%%%%%%%%%%%%%%%%
\section{The better theory?}
Which then of these theories is the better theory? Our viewpoint is that the
best is GR because it is the most fundamental of all these theories - indeed
it is {\it the} fundamental classical gravitational theory. The true theory
is GR. The issue is what are usable approximations in Newtonian-like
conditions (for example, in studying local astrophysics). \\

BH is a good candidate, but is certainly a generalisation of NT
rather than being classical NT. It reflects the causal structure of GR,
but is closer to linearised GR rather than full GR (this is no
accident; it was constructed that way). From this viewpoint, NG in turn
is an acceptable approximation to BH theory in some circumstances; but the
range of conditions adequately covered by BH theory is wider than that
of NG but less than that of GR.\\

Which is more useful in astrophysical contexts will depend on those contexts.
Thus we have not considered here the issue of which theories give adequate
description of anisotropic collapse situations leading to the formation
of structure in the expanding universe. However the line of argument above
suggests that where NT and BH theory disagree, we should believe the latter
(but where BH and GR disagree, we should believe GR). \\

Finally does Newtonian theory itself demand a non-zero magnetic Weyl
tensor analogue? No - it is probably not even well-defined. But that
fact does not undermine the claims of BH theory to give a better
description of what will happen in some collapse scenarios. But that does not
mean we should necessarily expect a non-zero Magnetic Weyl tensor (or its
quasi-Newtonian analogue) in realistic collapse situations. The point is
that one should treat carefully the claim that neighbouring world-lines do
not influence each other in `silent universe' [11], for they do indeed feel
each other's gravitational influence [24] through the set of constraint
equations -- including the relativistic version of the Poisson equation --
that are necessarily satisfied in a full solution of the field equations (and
are consistent with the evolution equations in such `silent' universes [25]).
It is gravitational induction and gravitational waves that are precluded
in these solutions - but these effects are not likely to be important
during the Newtonian-like phase of gravitational collapse.\\ \\
%%%%%%%%%%%%%%%%%%%%%%%%%%%%%%%%%%%%%%%%%%%%%%%%%%%%%%%%%%%%%%%%%%%%%%%%
We thank the FRD (South Africa) for financial support.\\ \\
%%%%%%%%%%%%%%%%%%%%%%%%%%%%%%%%%%%%%%%%%%%%%%%%%%%%%%%%%%%%%%%%%%%%%%%%

{\bf References}\\

[1] E Bertschinger and A J S Hamilton: Lagrangian Evolution of the Weyl
Tensor.  MIT preprint (1994) submitted to ApJ.

[2] W Kundt and M Trumper: Akad Wiss u Lit Mainz Math-Nat Kl, 12 (1961)

[3] S W Hawking: ApJ {\bf 145} 544 (1966)

[4] G F R Ellis: in General Relativity and Cosmology, ed. R K Sachs
(New York: Academic Press), 104 (1971)

[5] G F R Ellis: in Cargese Lectures in Physics, Vol 6, Ed E Schatzmann
(New York: Gordon and Breach), 1 (1973)

[6] G F R Ellis and M Bruni: ``A covariant and gauge-free approach to
density fluctuations in cosmology". {\em Phys Rev} {\bf D40}, 1804-1818
(1989).

[7] G F R Ellis, J Hwang, and M Bruni: ``Covariant and gauge-
independent perfect fluid Robertson-Walker perturbations". {\em Phys
Rev D.} 40, 1819-1826 (1989).

[8] G F R Ellis, M Bruni, and J C Hwang: ``Density-Gradient Vorticity
relation in perfect fluid Robertson-Walker perturbations". {\em Phys
Rev} {\bf D42} 1035-1046, (1990)

[9] M Bruni P K S Dunsby and G F R Ellis: ``Cosmological perturbations
and the physical meaning of gauge-invariant variables". {\em Astrophys
Journ} {\bf 395} 34-53 (1992)

[10] Matarrese, S., Pantano, O., and Saez, D. 1993,
{\it Phys. Rev.} D {\bf 47}, 1311.

[11] Matarrese, S., Pantano, O., and Saez, D. 1994,
{\it Phys. Rev. Lett.}, {\bf 72}, 320.

[12] K M Croudace, J Parry, D S Salopek, and J M Stewart, Applying the
Zel'dovich Approximation to general relativity. 1993.

[13] Bruni, M., Matarrese, S., and Pantano, O. Dynamics of
silent universes. 1994. Submitted to {\it Ap.J}.

[14] Bruni, M., Matarrese, S., and Pantano, O. 1994. A local view of
the observable universe. Submitted to {\it Ap.J}.

[15] Ya B Zeldovich, Ann Rev Fluid Mech 9:215-228 (1977)

[16] O Heckmann and E Schucking. Zs f Astrophysik 38:95-109 (1955)

[17] O Heckmann and E Schucking. Zs f Astrophysik 40:81-92 (1956)

[18] O Heckmann and E Schucking. Handbuch d Physik, Vol 53 (1959)

[19] J V Narlikar, Mon Not Roy Ast Soc 126: 203-208 (1963),
131:501-522 (1966)

[20] Ya B Zeldovich, Mon Not Roy Ast Soc 129: 19 (1965)

[21] G F R Ellis and G Schreiber: {\em Phys Lett} {\bf A115 },
97-107 (1986).

[22] G F R Ellis: In {\em General Relativity and Gravitation}, Ed B
Bertotti et al (Reidel, 1984), 215-288.

[23] G F R Ellis: {\em Mon Not Roy Ast Soc}, {\bf 243}, 509-516
(1990).

[24] G F R Ellis and D W Sciama: in {\em General Relativity}, ed. L.
O'Raifeartaigh (Oxford University Press, 1972), 35-59.

[25] W M Lesame, P K S Dunsby, and G F R Ellis, UCT preprint (1994).
\end{document}